\documentclass[12pt]{article}

\usepackage{amsfonts,amssymb,bm,cite,amsmath}
\usepackage[dvips]{graphicx}

\newcommand {\beq}{\begin{equation}}
\newcommand {\eeq}{\end{equation}}
\newcommand {\beqa}{\begin{eqnarray}}
\newcommand {\eeqa}{\end{eqnarray}}

\renewcommand{\theequation}{\thesection.\arabic{equation}}

\begin{document}
\setlength{\oddsidemargin}{0cm}
\setlength{\baselineskip}{7mm}

\begin{titlepage}
\renewcommand{\thefootnote}{\fnsymbol{footnote}}
\begin{normalsize}
\begin{flushright}
\begin{tabular}{l}
November 2015
\end{tabular}
\end{flushright}
  \end{normalsize}

~~\\

\vspace*{0cm}
    \begin{Large}
       \begin{center}
         {Entanglement entropy in scalar field theory on the fuzzy sphere}
       \end{center}
    \end{Large}
\vspace{1cm}

\begin{center}
           Shizuka O{\sc kuno}$^{1)}$\footnote
            {
e-mail address : 
okuno.shizuka.14@shizuoka.ac.jp},
           Mariko S{\sc uzuki}$^{1,2)}$\footnote
            {
e-mail address : 
f5344003@ipc.shizuoka.ac.jp}
           {\sc and}
           Asato T{\sc suchiya}$^{1,2)}$\footnote
           {
e-mail address : tsuchiya.asato@shizuoka.ac.jp}\\
      \vspace{1cm}

 $^{1)}$ {\it Department of Physics, Shizuoka University}\\
                {\it 836 Ohya, Suruga-ku, Shizuoka 422-8529, Japan}\\
         \vspace{0.3cm}     
        $^{2)}$ {\it Graduate School of Science and Technology, Shizuoka University}\\
               {\it 3-5-1 Johoku, Naka-ku, Hamamatsu 432-8011, Japan}

\end{center}

\hspace{5cm}

\begin{abstract}
\noindent
We study entanglement entropy on the fuzzy sphere.
We calculate it in a scalar field theory on the fuzzy sphere,
which is given by a matrix model. 
We use a method that is based on the replica method and applicable to
interacting fields as well as free fields.
For free fields, we obtain the results consistent with the previous study,
which serves as a test of the validity of the method.
For interacting fields,
we perform Monte Carlo simulations at strong coupling
and see a novel behavior of entanglement entropy.
\end{abstract}
\vfill
\end{titlepage}
\vfil\eject

\setcounter{footnote}{0}

\section{Introduction}
\setcounter{equation}{0}
\renewcommand{\thefootnote}{\arabic{footnote}} 
Since the discovery of the Ryu-Takayanagi formula \cite{Ryu:2006bv}
in the context of
the AdS/CFT correspondence, it has been
revealed that entanglement entropy in field theory
encodes the information on geometry.
Because the noncommutative field theory is deeply connected to
gravity and string theory, it would be worthwhile to study entanglement entropy
in such field theories.
Indeed, by examining  a gravity dual of noncommutative Yang-Mills theory 
proposed in \cite{Hashimoto:1999ut,Maldacena:1999mh}, 
it was conjectured in \cite{Fischler:2013gsa,Karczmarek:2013xxa} that 
the leading contribution to entanglement entropy in noncommutative 
Yang-Mills theory is proportional  to the volume of the focused region\footnote{
In this paper, we use terminologies `volume' and `area' even for 
sphere, although their actual meanings are area and length on sphere, respectively.}, 
in contrast to the fact
that it is proportional to the area of the boundary in ordinary field theories.
This volume law is considered to originate 
from the UV/IR mixing \cite{Minwalla:1999px} due to
nonlocal nature of interactions.
In fact,  in \cite{Shiba:2013jja}, the volume law is observed in nonlocal theories.

In \cite{Karczmarek:2013jca,Sabella-Garnier:2014fda}\footnote{See also 
\cite{Dou:2006ni,Dou:2009cw}.}, 
entanglement entropy in a scalar field theory
on the fuzzy sphere, 
which is realized by a matrix model,
was calculated for free fields\footnote{In this paper, we  call
the case in which the action consists of only quadratic terms `free field' while 
the case in which the action includes higher terms  `interacting 
field'.}. 
In this paper, we are concerned with interacting fields, because
the discovered UV/IR anomaly \cite{Chu:2001xi,CastroVillarreal:2004vh} 
that is a counterpart of the UV/IR mixing in 
field theories on compact noncommutative manifolds arises from the interactions.

Another motivation of our work is to gain insights into connections between
geometry and matrix models, which is an important subject 
in the context of matrix models
proposed as nonperturbative formulation of string theory \cite{BFSS,IKKT,DVV}.
 In \cite{Karczmarek:2013jca,Sabella-Garnier:2014fda},
the fuzzy sphere is divided into two regions using the Bloch coherent state, and
correspondingly the matrices are divided into two parts.
By considering the results for entanglement entropy, 
we should elucidate a precise geometrical meaning of this division.

In \cite{Karczmarek:2013jca,Sabella-Garnier:2014fda}, 
the method developed 
in \cite{Srednicki:1993im} was used to calculate entanglement entropy.
This method is valid only for free fields.
In this paper, we use another method, which was developed and used
in \cite{Buividovich:2008kq,Nakagawa:2011su} and
can also be applied to interacting fields. We first test the validity of the method
in our study
by applying the method to free fields. We compare our results with those 
in \cite{Karczmarek:2013jca,Sabella-Garnier:2014fda}. Then, we apply the method
to interacting fields and perform Monte Carlo simulations\footnote{For Monte Carlo
simulations of the fuzzy sphere, see \cite{Azuma:2004zq,Azuma:2005bj,
Medina:2005su,GarciaFlores:2009hf,Panero:2006bx,Das:2007gm}. For analytic treatment
of scalar field theory on the fuzzy sphere, 
see \cite{Steinacker:2003sd,Kawamoto:2015qla}.}. 
While this work is a first step to Monte Carlo study of entanglement entropy on
the fuzzy sphere, we see a novel behavior of
entanglement entropy.

This paper is organized as follows.
In section 2, we review a matrix model that realizes
a noncommutative counterpart of a scalar field theory on $S^1\times S^2$.
In section 3, we describe the properties of
entanglement entropy and explain the method to calculate entanglement
entropy. In section 4, we show numerical results for entanglement entropy.
Section 5 is devoted to discussion. In appendix A, we review the Bloch coherent state.
In appendix B, the detail of the calculation in the free case is given.

\section{Scalar field theory on the fuzzy sphere}
\setcounter{equation}{0}
First, we consider a scalar field theory on $S^1\times S^2$ defined by
\begin{align}
S_C=\frac{R^2}{4\pi}\int_0^{\beta} dt \int d\Omega \left
(\frac{1}{2}\dot{\phi}^2-\frac{1}{2R^2}({\cal L}_i \phi)^2+\frac{\mu^2}{2}\phi^2
+\frac{\lambda}{4}\phi^4 \right) \ ,
\label{commutative action}
\end{align}
where $\beta$ is the circumference of $S^1$ that corresponds to inverse temperature, 
$R$ is the radius of $S^2$,
the integral measure on $S^2$
is given by 
$R^2\int d\Omega=R^2\int_0^{2\pi} d\varphi \int_0^{\pi} d\theta \sin\theta$, and
the dot stands for the derivative with respect to $t$.
${\cal L}_i$ (i=1,2,3) are the orbital angular momentum operators that are defined by
\begin{align}
{\cal L}_{\pm} &\equiv {\cal L}_1 \pm i {\cal L}_2 
=e^{\pm i\varphi} \left( \pm \frac{\partial}{\partial \theta} 
+i \cot \theta \frac{\partial}{\partial \varphi} \right) \ ,\nonumber\\
{\cal L}_3 &= -i \frac{\partial}{\partial \varphi} \ .
\end{align}

A noncommutative counterpart of the theory (\ref{commutative action}), where
$S^2$ is replaced with the fuzzy sphere, is given by
a matrix model, whose action is defined by
\begin{align}
S_{NC}=\frac{R^2}{2j+1}\int_0^{\beta} dt \  \mbox{tr} \left (\frac{1}{2}\dot{\Phi}^2
-\frac{1}{2R^2}[L_i,\Phi]^2 +\frac{\mu^2}{2}\Phi^2+\frac{\lambda}{4}\Phi^4 \right) \ ,
\label{noncommutative action}
\end{align}
where $j$ is a non-negative integer or a positive half-integer, and  
$\Phi$ is a $(2j+1)\times (2j+1)$ Hermitian
matrix depending on $t$. $L_i$ are the generators of the $SU(2)$ algebra
with the spin $j$ representation
obeying the relation
$ [L_i, L_j]=i\epsilon_{ijk} L_k $.
The theory (\ref{noncommutative action}) reduces 
to the theory (\ref{commutative action}) at the tree level in the limit $j\rightarrow \infty$,
which corresponds to the continuum limit,
while the theory (\ref{noncommutative action}) exhibits the UV/IR anomaly 
at the quantum level, 
which makes the theory (\ref{noncommutative action}) differ from
the theory (\ref{commutative action}) even in the $j\rightarrow \infty$ limit.

A simple way to see the correspondence between the two theories is to use
the Bloch coherent 
states $|\Omega \rangle$ $(\Omega=(\theta,\varphi)$) \cite{Gazeau}\footnote{See also
\cite{Alexanian:2000uz,Hammou:2001cc,Presnajder:1999ky,Ishiki:2015saa}.}, 
which are reviewed in appendix A.
We identify the Berezin symbol 
$f_{\Phi(t)}(\Omega)=\langle\Omega|\Phi(t)|\Omega\rangle$ \cite{Berezin:1974du}
with $\phi(t,\Omega)$ in the $j\rightarrow\infty$ limit.
By using (\ref{explicit form}), one can easily show that 
\begin{align}
f_{[L_i,\Phi] }(\Omega)={\cal L}_i f_{\Phi}(\Omega) \ .
\end{align}
Moreover, for two matrices $A$ and $B$, the star product is given by
\begin{align}
f_{A}(\Omega)\ast f_{B}(\Omega) \equiv f_{AB}(\Omega)
=\frac{2j+1}{4\pi}\int d\Omega'  \ \langle\Omega | A | \Omega'\rangle
\langle\Omega'|B|\Omega\rangle \ ,
\end{align}
where we used (\ref{property4}). The star product reduces to the ordinary
product at the tree level in the $j\rightarrow\infty$ limit, while it yields the UV/IR anomaly at the quantum level. Thus the theory (\ref{noncommutative action}) reduces to 
the theory (\ref{commutative action}) at the tree level in the $j\rightarrow\infty$ limit.

The relationship between the Berezin symbol and 
the matrix elements $\langle jm | \Phi |jm'\rangle$ is given by
\begin{align}
f_{\Phi}(\Omega)=\sum_{m,m'} \langle\Omega | jm\rangle \langle jm' | \Omega\rangle
\langle jm | \Phi | jm'\rangle  \ .
\end{align}
Here, by  using (\ref{explicit form}), one finds that
\begin{align}
\langle\Omega | jm\rangle \langle jm' | \Omega\rangle
\sim \left(\cos\frac{\theta}{2}\right)^{2j+m+m'}\left(\sin\frac{\theta}{2}\right)^{2j-m-m'}
e^{i(m-m')\varphi}  \ ,
\end{align}
which turns out to have the sharp peak at \cite{Karczmarek:2013jca}
\begin{align}
\cos\theta = \frac{m+m'}{2j} \ .
\label{region on sphere and matrix}
\end{align}
The width is given by $\Delta \theta \sim \frac{1}{\sqrt{j}}$.
This implies that the matrix elements  $\langle jm | \Phi | j \ n-m\rangle$ 
correspond to the field $\phi$ at $\cos\theta=\frac{n}{2j}$ \cite{Karczmarek:2013jca}.

Hereafter, we put $R=1$ and $\mu=1$ for simplicity,
and denote the matrix size by $N$, namely $N=2j+1$.



\section{Calculation of entanglement entropy}
\setcounter{equation}{0}
In this section, we first review the properties of entanglement entropy and
next explain how to calculate entanglement entropy on the fuzzy sphere.
\subsection{Entanglement entropy}
\begin{figure}[tbp]
\begin{center}
\begin{minipage}{0.2\hsize}
 \begin{center}
  {\includegraphics[width=50mm]{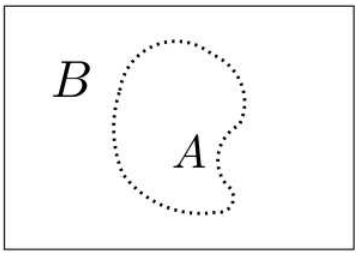}}
 \end{center}
\end{minipage}
\end{center}
\caption{Region A and region B.}
\label{divide}
\end{figure}

Suppose that the Hilbert space ${\cal H}$ of a system is given by
a tensor product 
\begin{align}
{\cal H}={\cal H}_A\otimes {\cal H}_B \ .
\label{decomposition of Hilbert space}
\end{align}
Then, the entanglement entropy $S_A$ is defined by
\begin{align}
S_{A}=- \mathrm{Tr} [\rho _{A}\log \rho _{A}] \ .
\label{definition of entanglement entropy}
\end{align}
Here $\rho_A$ is obtained by taking a partial trace of the density matrix $\rho_{tot}$
over ${\cal H}_B$:
\begin{align}
\rho_A = \mathrm{Tr}_B[\rho_{tot}] \ .
\end{align}
Typically, the decomposition of the Hilbert space
(\ref{decomposition of Hilbert space}) is realized by a decomposition of the space,
on which a field theory is defined,
into two regions, as the region A and the region B  in Fig.\ref{divide}.

Entanglement entropy has the following properties. 
First, if the density matrix $\rho_{tot}$ is given by a pure state,
entanglement entropy satisfies
\begin{align}
S_A=S_B \ .
\label{rhoA=rhoB}
\end{align}
Second, 
the leading contribution to the entanglement
entropy $S_A$ for the ground state
in $(d+1)$-dimensional local field theories ($d\geq 2$)
is proportional to $|\partial A|/\epsilon^{d-1}$, where
$|\partial A|$ is the area of the boundary between the regions A and B, and
$\epsilon$ is the UV cutoff.
At finite temperature, entanglement entropy has a correction proportional to
the volume of the region A.
On the other hand,  in nonlocal field theories,
the leading contribution to the entanglement entropy $S_A$
for the ground state can
be proportional to the volume of the region A.
In particular, by examining the gravity dual, it was conjectured in 
\cite{Fischler:2013gsa,Karczmarek:2013xxa} that
this is indeed the case in noncommutative Yang-Mills theory.



\begin{figure}[tbp]
\begin{center}
{\includegraphics[width=100mm]{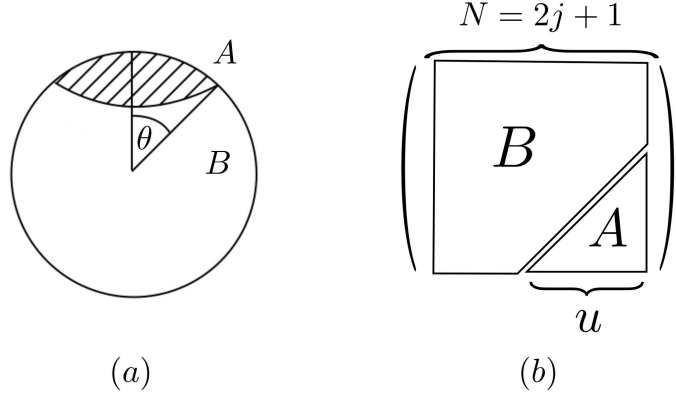}}
\end{center}
\label{sphere}
\caption{Correspondence of two regions on the fuzzy sphere and in the matrix model.}
\end{figure}

In our study, we divide the fuzzy sphere into two region, as in \cite{Karczmarek:2013jca}.
By using (\ref{region on sphere and matrix}), we identify the regions A and B
on the sphere in Fig.\ref{sphere}(a) with the regions A and B 
of the matrix $\Phi$
in Fig.\ref{sphere}(b), respectively.
In order to specify the regions A and B on the sphere, 
we introduce a new parameter $x$, which is related to $\theta$ as
\beq
x=1-\cos \theta \ .
\label{x and theta}
\eeq
Namely, $x$ is the area of the region A divided by $2\pi$.
The condition that the $(m,m')$ component of the matrix $\Phi$ is located 
in the region A is given by
\begin{align}
m+m' > 2j-u \ ,
\label{m+m'}
\end{align}
where $u=0,1,2,\cdots,4j$.
Then, it follows from  (\ref{region on sphere and matrix}), (\ref{x and theta}) 
and (\ref{m+m'}) that
the relation between $x$ and $u$ is given by
\begin{align}
x=\frac{u}{2j} \ .
\label{relation between x and u}
\end{align}

\subsection{Replica method}
In this subsection, we describe the method to calculate entanglement entropy 
developed in \cite{Buividovich:2008kq}.
In calculating entanglement entropy,  we use  the replica method,
in which the definition of entanglement entropy 
 (\ref{definition of entanglement entropy}) is rewritten as
\beq
S_{A}=\lim _{\alpha\rightarrow 1}\left[-\frac{\partial}{\partial \alpha}\mathrm{Tr}\rho_{A}^{\alpha}\right]=\lim _{\alpha\rightarrow 1}\left[-\frac{\partial}{\partial \alpha}\log (\mathrm{Tr}\rho_{A}^{\alpha})\right] \ ,
\label{difS}
\eeq
where $\alpha$ corresponds to the number of replicas and is analytically continued.

\begin{figure}[t]
 \begin{center}
  {\includegraphics[width=50mm]{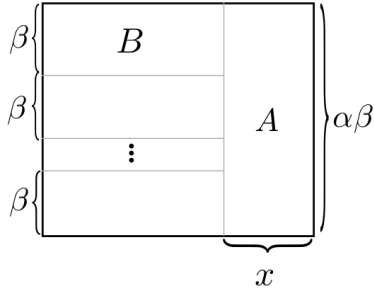}}
 \end{center}
\caption{Replica method.}
\label{fig:replica}
\end{figure}

In (\ref{noncommutative action}), we yield $\alpha$ replicas for $\Phi(t)$, 
which are denoted
by $\Phi_n(t)$ $(n=1,\cdots,\alpha)$. 
We impose the following boundary condition on $\Phi_n(t)$ (see Fig.\ref{fig:replica}):
\begin{align}
\Phi_n(\beta,m,m')&=\Phi_{n+1}(0,m,m')  \;\; \mbox{for the region A}  \ , \nonumber\\
\Phi_n(\beta,m,m')&=\Phi_n(0,m,m') \;\; \mbox{for the region B} \ ,
\end{align}
where $n=1,\cdots,\alpha$ and $\alpha+1$ is identified with 1 in the first line.
Then, we obtain a relation
\beq
\mathrm{Tr}\rho^{\alpha}_{A}=\frac{Z(x,\alpha)}{Z^\alpha} \ ,
\label{rho}
\eeq
where $Z$ represents $Z(\alpha=1)$ that is independent of $x$. 
Substituting (\ref{rho}) into (\ref{difS})
leads to an expression for $S_A$
\beq
S_A(x)=-\lim_{\alpha\rightarrow 1}\frac{\partial}{\partial \alpha} \ln\left(\frac{Z(x,\alpha)}{Z^\alpha}\right) \ .
\eeq
We obtain entanglement entropy for the ground state 
in the $\beta\rightarrow\infty$ limit, while one including finite temperature effect
at finite $\beta$.

It is convenient to consider the derivative of $S_A$ with respect to $x$ 
instead of $S_A$ itself:
\beq
\frac{\partial S_A(x)}{\partial x}
=\frac{\partial}{\partial x}\left[-\lim_{\alpha\rightarrow 1}\frac{\partial}{\partial \alpha}\ln\left(\frac{Z(x,\alpha)}{Z^\alpha}\right)\right]=\lim _{\alpha\rightarrow 1}\frac{\partial}{\partial x}\frac{\partial}{\partial \alpha}F[x,\alpha] \ ,
\eeq
where $F[x,\alpha]$ is the free energy of the system Fig. \ref{fig:replica}. 
Here we make an approximation\footnote{
Precisely speaking, we calculate the derivative of 
the R\'enyi entropy with the R\'enyi parameter
equal to two with respect to $x$.} for the derivative with respect to $\alpha$ as
\begin{align}
&\lim _{\alpha\rightarrow 1}
\frac{\partial}{\partial x}\frac{\partial}{\partial \alpha}F[x,\alpha] \nonumber\\
&\rightarrow 
\frac{\partial}{\partial x}(F[x,\alpha=2]-F[x,\alpha=1]) 
=\lim_{j\rightarrow\infty} \frac{F[x+\varepsilon,\alpha=2]-F[x,\alpha=2]}{\varepsilon}
\ ,
\label{approximation}
\end{align}
where $\varepsilon=\frac{1}{2j}$.
In the next section, we test the validity of this approximation 
by comparing our results for 
free fields
with those in \cite{Karczmarek:2013jca,Sabella-Garnier:2014fda}.
Note that (\ref{rhoA=rhoB}) implies that in the $\beta\rightarrow\infty$ limit
\begin{align}
&S_A(x)=S_A(2-x) \ , \nonumber\\
&\frac{\partial S_A}{\partial x}(x)=-\frac{\partial S_A}{\partial x}(2-x) \ ,
\label{odd function}
\end{align}
which reflect the symmetry under $\theta \rightarrow \pi-\theta$.

In the case of free fields  where $\lambda=0$, we calculate $F[x,\alpha=2]$
directly by evaluating numerically the determinant that is
given in appendix B.

In the case of interacting fields where $\lambda\neq 0$, it is convenient to
introduce an interpolating action
$S_{int}=(1-\gamma)S_{x+\varepsilon}+\gamma S_{x}$, where
$S_{x+\varepsilon}$ and $S_{x}$ are the actions that would yield
$F[x+\varepsilon,\alpha=2]$ and $F[x,\alpha=2]$, respectively.
Then the numerator of the last expression in 
(\ref{approximation}) 
can be evaluated as
\beq
F[x+\varepsilon,\alpha=2]-F[x,\alpha=2]
=\int_0 ^1 d\gamma  \ \langle S_{x+\varepsilon}-S_{x}\rangle _\gamma \ ,
\label{difF}
\eeq
where $\langle \cdots \rangle_\gamma$ stands for the expectation value
with respect to the canonical weight $e^{-S_{int}}$.
In practice, we take $\gamma$ from 0 to 1 by the step 0.1, 
and calculate $\langle S_{x+\varepsilon}-S_{x}\rangle _\gamma$ for each $\gamma$.
We finally use the Simpson formula for the integral to obtain the right-hand side of
(\ref{difF}).

In both cases, we introduce the lattice in the time direction 
and denote the lattice spacing
by $a$.

\section{Results}
\setcounter{equation}{0}
In this section, we show our results for free fields ($\lambda=0$) 
and for interacting fields. In the latter case, we put $\lambda=1.0$,
which would correspond to a strong coupling.

\subsection{$\lambda=0$}
\begin{figure}[t]
\begin{center}
 \begin{center}
  {\includegraphics[width=110mm]{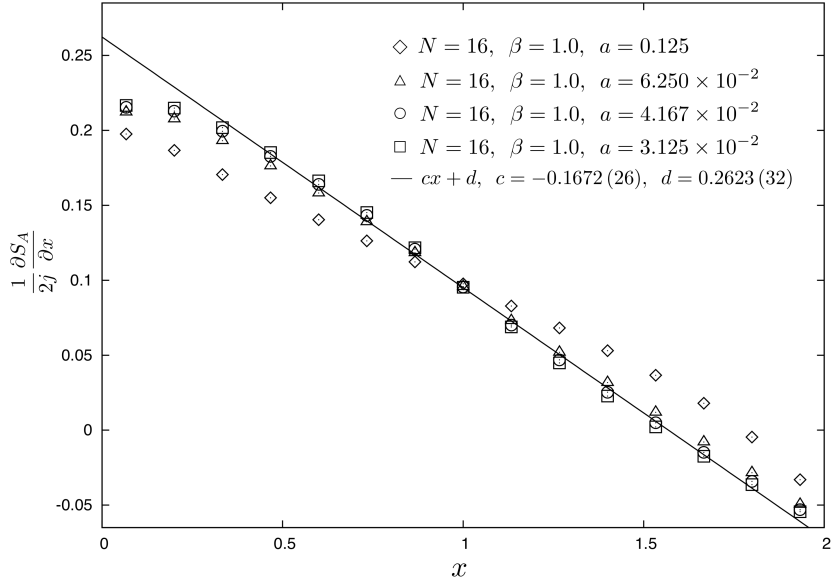}}
 \end{center}
\end{center}
\caption{The derivative of  entanglement entropy with respect to $x$ divided by
$2j$ is plotted against $x$ at $\lambda=0$, $N=16$ and $\beta=1.0$.
The diamonds, the triangles, the circles and the squares represent the data for
$a=0.125, 6.250\times 10^{-2}, 4.167\times 10^{-2}, 3.125\times 10^{-2}$, 
respectively. The solid line is a fit of the data for $a=3.125\times 10^{-2}$ to
$\frac{1}{2j}\frac{\partial S_A}{\partial x}=cx+d$ for
$0.333\leq x \leq 1.800$, which gives 
$c=-0.1672(26)$ and $d=0.2623(32)$. }
\label{beta=1}
\end{figure}

\begin{figure}[t]
\begin{center}
 \begin{center}
  {\includegraphics[width=110mm]{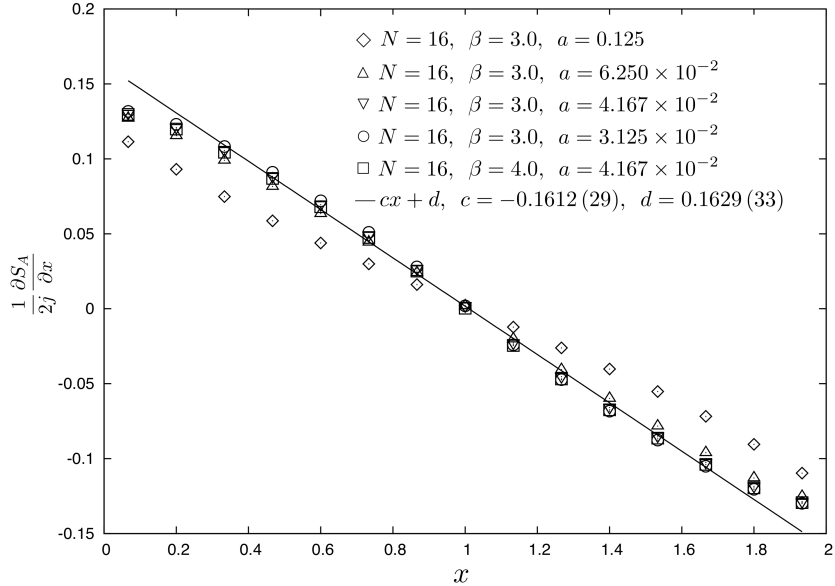}}
 \end{center}
\end{center}
\caption{The derivative of  entanglement entropy with respect to $x$ divided by
$2j$ is plotted against $x$ at $\lambda=0$ and $N=16$.
The diamonds, the triangles, the inverted triangles and the circles
represent the data for $\beta=3.0$ and
$a=0.125, 6.250\times 10^{-2}, 4.167\times 10^{-2}, 3.125\times 10^{-2}$, respectively,
while the squares represent the data for $\beta=4.0$ and
$a=4.167\times 10^{-2}$.
The solid line is a fit of the data for $\beta=3.0$ and $a=3.125\times 10^{-2}$ to
$\frac{1}{2j}\frac{\partial S_A}{\partial x}=cx+d$ for
$0.200\leq x \leq 1.800$, which gives 
$c=-0.1612(29)$ and $d=0.1629(33)$. }
\label{beta=3,4}
\end{figure}

\begin{figure}[t]
\begin{center}
 \begin{center}
  {\includegraphics[width=110mm]{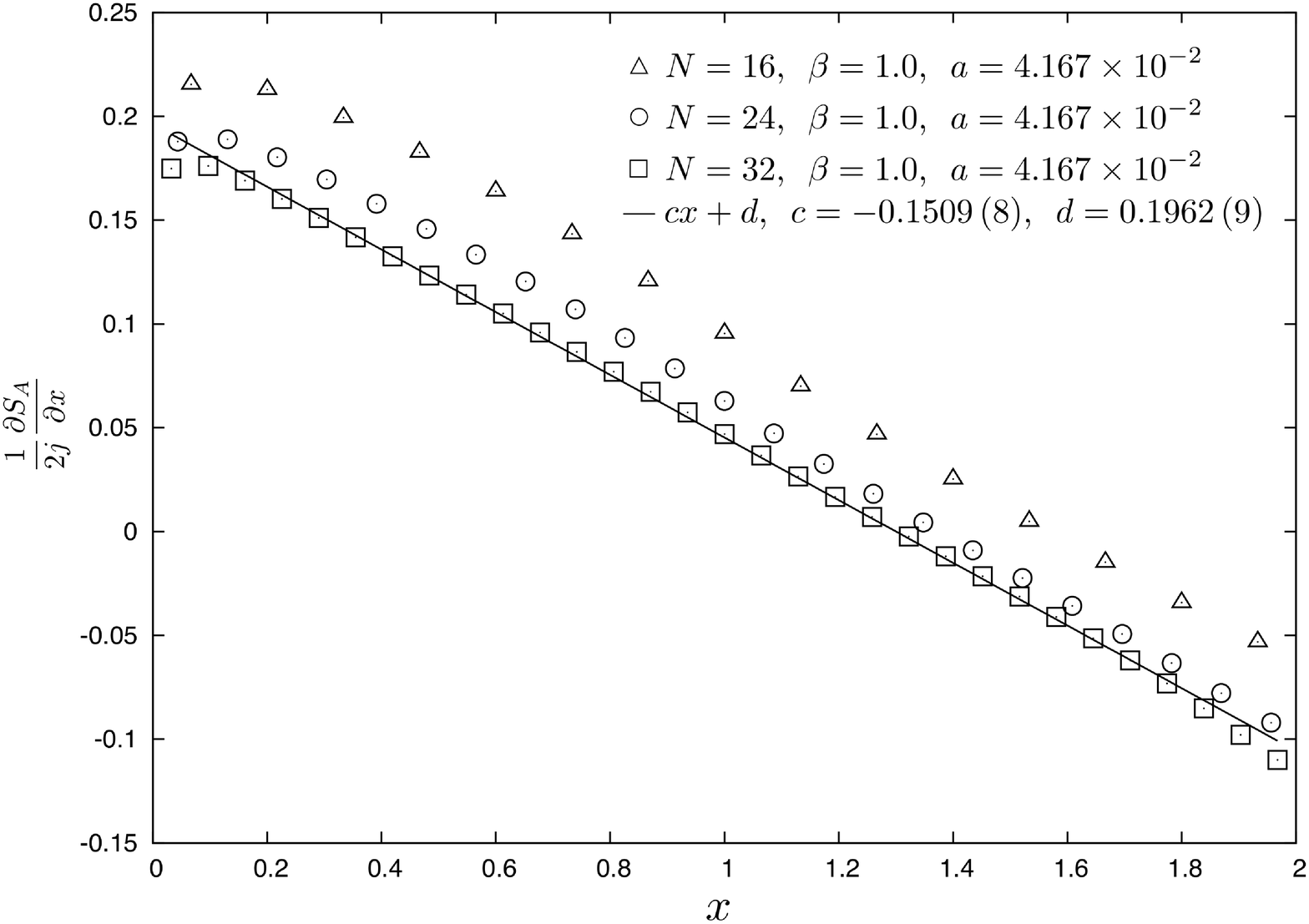}}
 \end{center}
\end{center}
\caption{The derivative of entanglement entropy with respect to $x$ divided by
$2j$ is plotted against $x$ at $\lambda=0$, $\beta=1.0$ and
$a=4.167\times 10^{-2}$.
The triangles, the circles and the squares
represent the data for $N=16, 24, 32$, respectively.
The solid line is a fit of the data for $N=32$ to
$\frac{1}{2j}\frac{\partial S_A}{\partial x}=cx+d$ for
$9.677\times 10^{-2}\leq x \leq 1.903$, which gives 
$c=-0.1509(8)$ and $d=0.1962(9)$.}
\label{manyN-line}
\end{figure}

We first calculate $F[x,\alpha=2]$ numerically by the method given in appendix B
and then calculate the derivative of the entanglement entropy $S_A$ with respect to $x$
following (\ref{approximation}).
The derivative of $S_A$ with respect to $x$ divided by $2j$ is plotted against $x$ 
in Fig. \ref{beta=1} - Fig. \ref{manyN-line}. 



We observe that at $\beta=1.0$ the data for odd $u$ exhibits a smooth behavior
while the data for even $u$ exhibits another smooth behavior 
(note that $x=\frac{u}{2j}$).  This discrepancy almost 
disappears at $\beta=4.0$. This discrepancy is considered to come
from a finite $N$ effect that becomes stronger at high temperature. 
Indeed, as we will see shortly, the continuum limit in the
time direction can be taken at $\beta=1.0$ using only the data for odd $u$ or 
even $u$ (see Fig. \ref{beta=1} for odd $u$ at $\beta=1.0$ and Fig. \ref{beta=3,4} for
odd $u$ at $\beta=3.0$), so that the two continuum limits for odd $u$ and  for 
even $u$ differ only by finite temperature effect.
Because we are concerned with the part except the finite temperature effect,
we plot only the date
for odd $u$ in the following.


In Fig. \ref{beta=1} and Fig. \ref{beta=3,4}, 
we examine the continuum limit in the time direction at $N=16$
and $\beta=1.0$ and at $N=16$ and $\beta=3.0$, respectively. 
We plot the data for four different values of the lattice spacing $a$. 
We observe that the continuum limit is taken, and $a=4.167\times 10^{-2}$ is 
close enough to the continuum limit.
The data for $a=3.125 \times 10^{-2}$ is fitted to the linear function
$\frac{1}{2j}\frac{\partial S_A}{\partial x}=cx+d$, where
we exclude some data points around $x=0$ and $x=2.0$, where the area of 
the region A or the region B is small so that 
ambiguity of the boundary between the two regions due to finite $N$ effect  is relevant.
We use the range $0.333\leq x \leq 1.8$ for $\beta=1.0$ and 
the range $0.2 \leq x \leq 1.8$ for $\beta=3.0$.
We obtain $c=-0.1672(26)$ and $d=0.2623(32)$ for $\beta=1.0$
and $c=-0.1612(29)$ and $d=0.1629(33)$ for $\beta=3.0$.

In Fig. \ref{beta=3,4}, 
we also plot the data for $N=16$, $\beta=4.0$ and 
$a=4.167\times 10^{-2}$.  We see that the data almost agree with those 
for $N=16$, $\beta=3.0$ and $a=4.167 \times 10^{-2}$. 
This implies that the low temperature limit (the $\beta\rightarrow\infty$ limit)
is taken and that  $\beta=3.0$ is close enough to the low temperature limit.
Indeed, the function $\frac{1}{2j}\frac{\partial S_A}{\partial x}=cx+d$ with $c=-0.1612(29)$ and $d=0.1629(33)$ to which
the data for $N=16$, $\beta=3.0$ and $a=3.125 \times 10^{-2}$ are fitted 
is consistent with
(\ref{odd function}). Namely, the function is proportional to $1-x$ within the 
fitting error. This implies that
\begin{align}
S_A \propto 2x-x^2= \sin^2\theta \ .
\label{sin^2theta}
\end{align}
This behavior agrees with the one  observed 
in \cite{Karczmarek:2013jca,Sabella-Garnier:2014fda} up to
an overall coefficient.

By comparing the above values of $c$ and $d$ obtained in the fitting of the data
for $\beta=1.0$ with those obtained in the fitting of 
the data for $\beta=3.0$,  we see that
the difference of the two functions $\frac{1}{2j}\frac{\partial S_A}{\partial x}=cx+d$ 
is almost constant.  This implies that the finite temperature contribution 
to entanglement entropy is proportional to $x$, namely the volume 
of the region A. This is a general property of entanglement entropy.
We also fit the data with even $u$ for $N=16$, $\beta=1.0$ and
$a=3.125 \times 10^{-2}$ to $\frac{1}{2j}\frac{\partial S_A}{\partial x}=cx+d$
for $0.133 \leq x \leq 1.6$ and obtain
$c=0.1626(26)$ and $d=0.2690(22)$. As we stated, 
the difference between the fitting of 
the data with odd $u$ and the one of the data with even $u$ is almost constant,
which is finite temperature effect.

In Fig. \ref{manyN-line}, we examine the large-$N$ (large-$j$) limit.
At $\beta=1.0$ and $a=4.167\times 10^{-2}$,
we plot the data for $N=16, 24, 32$. We observe that the data converge as $N$
increases. This implies that entanglement entropy scales as $N$, which is 
consistent with the observation in \cite{Karczmarek:2013jca,Sabella-Garnier:2014fda}. 
We fit the data for $N=32$ to the function 
$\frac{1}{2j}\frac{\partial S_A}{\partial x}=cx+d$ for $0.0967 \leq x \leq 1.903$ and 
obtain 
$c=-0.1509(8)$ and $d=0.1962(9)$.
Thus, our method is valid in the sense that it reproduces 
the $\theta$ dependence (\ref{sin^2theta})
and the $N$ dependence of entanglement entropy precisely.

\subsection{$\lambda=1.0$}
In this subsection, we study the case of $\lambda=1.0$.
In the previous subsection,  we saw in the case of $\lambda=0$ that
the finite temperature effect is controllable.
Thus, as a first step,
we decide to perform Monte Carlo simulations at $N=16$, $\beta=1.0$ and
$a=0.125$ taking into account the computation time.

We use the Hybrid Monte Carlo method and make $3,000,000$ trajectories
for each $\gamma=0.0,0.1,\cdots,1.0$, 
discarding the first $100,000$ trajectories for the thermalization.

\begin{figure}[t]
\begin{center}
 \begin{center}
  {\includegraphics[width=110mm]{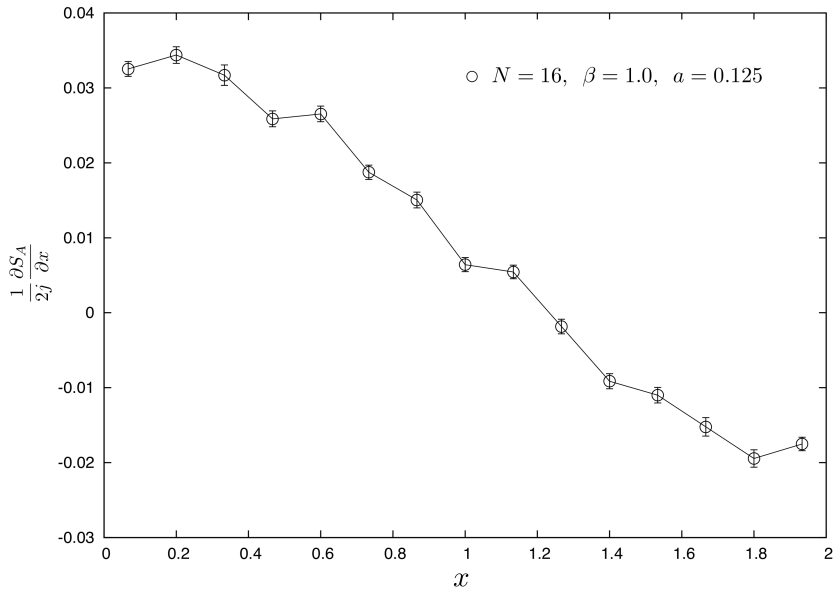}}
 \end{center}
\end{center}
\caption{The derivative of entanglement entropy with respect to $x$ divided by
$2j$ is plotted against $x$ at $\lambda=1.0$, $N=16$, $\beta=1.0$ and $a=0.125$.
}
\label{lambda=1}
\end{figure}

In Fig. \ref{lambda=1}, we plot the derivative of entanglement entropy with 
respect to $x$ divided by $2j$ against $x$. 
We again observe the discrepancy between odd $u$ and even $u$ 
similar to the case of
$\lambda=0$, so that we plot only the data for odd $u$.
We see that the data can be shifted
by a constant in the vertical
direction in such a way that they are consistent with (\ref{odd function})
except $u=13,15,17$.
Thus,  we conjecture that also 
in the case of interacting fields the finite temperature effect 
in entanglement
entropy is also proportional to the volume of the region A as in the  case of free fields.
Comparing Fig. \ref{lambda=1}  with Fig. \ref{beta=1}, 
we also see that the data for $\lambda=1.0$
behave in a clearly different way from the data for $\lambda=0$
with the same values of $N$, $\beta$ and $a$. 
Indeed, while the data for $\lambda=0$ can be fitted to
$\frac{1}{2j}\frac{\partial S_A}{\partial x}=cx+d$ with $c=-0.1276(33)$ and 
$d=0.2140(37)$
for $0.2\leq x \leq 1.933$, while 
the data for $\lambda=1.0$ cannot be fitted to such a linear function.
Furthermore, the magnitude of entanglement entropy 
for $\lambda=0$ is 
about ten times larger than that for $\lambda=1.0$.
We conjecture that this drastic difference is attributed to nonlocal
interactions as well as strong coupling.

\section{Discussion}
\setcounter{equation}{0}
In this paper, we calculated entanglement entropy in the scalar field theory
on the fuzzy sphere. We use the method developed and used 
in \cite{Buividovich:2008kq,Nakagawa:2011su}.
In the case of $\lambda=0$, we obtained the results that are consistent with
those in \cite{Karczmarek:2013jca,Sabella-Garnier:2014fda}. 
This serves as a check of the validity of the method in our study.
We performed  Monte Carlo simulations to calculate entanglement entropy 
at strong coupling ($\lambda=1.0)$. This is the first result for interacting fields
on the fuzzy sphere.

We found in the case of free fields that the finite temperature effect in 
entanglement entropy is proportional to the volume of the focused region as
in ordinary field theories.
We conjecture
from the result of Monte Carlo simulations that 
the same is true for the case of interacting fields.
We saw that the behavior of   entanglement entropy for interacting fields is 
clearly different from that for free fields. 
In particular, we found that magnitude of 
entanglement entropy for free free fields is about ten times larger than
that for interacting fields.
We conjecture that this drastic difference is attributed to nonlocal interactions
as well as strong coupling.

For free fields, we confirmed the observation 
in \cite{Karczmarek:2013jca,Sabella-Garnier:2014fda} that 
the entanglement entropy for the ground state is proportional to 
the square of the area of the boundary ($\sim \sin^2\theta$) and scales as $N$.
For interacting fields, we should examine the continuum limit and establish
the $\theta$ dependence of entanglement entropy, which is naively 
expected to be proportional to the volume.
We should give a physical interpretation on the behavior 
of entanglement entropy for interacting fields
as well as for free fields.
We would also like to study the $\lambda$ dependence of entanglement entropy. 
In particular, we are interested in whether there exists a phase transition or not.
By continuing Monte Carlo simulations, we hope to report on
the above issues in the
near future.


\section*{Acknowledgements}
We would like to thank G. Ishiki and E. Itou for discussions.
Numerical computation was carried out on SR16000 at YITP in Kyoto University
and SR16000 at University of Tokyo.
The work of A.T. is supported in part by Grant-in-Aid
for Scientific Research
(No. 24540264, 23244057 and 15K05046)
from JSPS.


\section*{Appendix A: Bloch coherent states}
\setcounter{equation}{0}
\renewcommand{\theequation}{A.\arabic{equation}}
In this appendix, we review the Bloch coherent state \cite{Gazeau}.
We introduce a standard basis $|jm\rangle$ $(m=-j, -j+1,\cdots,j)$
for the spin $j$ representation of the $SU(2)$ algebra,
which obey the relations
\begin{align}
L_{\pm}|jm\rangle &=\sqrt{(j\mp m)(j \pm m+1)}|j m\pm 1\rangle, \nonumber\\
L_3|jm\rangle &= m |jm\rangle \ ,
\end{align}
where $L_{\pm}=L_1 \pm i L_2$.
We consider the state $|jj\rangle$ to correspond to the north pole on unit sphere.
Then, the state $|\Omega\rangle$ 
that corresponds to a point $\Omega=(\theta,\varphi)$ on unit sphere is obtained
by multiplying $|jj\rangle$ by a rotation operator:
\begin{align}
|\Omega\rangle=e^{i\theta (\sin\varphi L_1 -\cos\varphi L_2)}|jj\rangle \ ,
\label{definition of coherent state}
\end{align}
from which it follows that
\begin{align}
n_iL_i |\Omega\rangle =j |\Omega\rangle  \ ,
\label{property 1}
\end{align}
where $\vec{n}=(\sin\theta\cos\varphi,\sin\theta\sin\varphi,\cos\theta)$.
This implies that the states $|\Omega\rangle$ 
minimize $\sum_i (\Delta L_i)^2$, where $(\Delta L_i)^2$
is the standard deviation of $L_i$.
The states $|\Omega\rangle$ are called the Bloch coherent states. 
(\ref{definition of coherent state}) is rewritten as
\begin{align}
|\Omega\rangle=e^{zL_-} e^{-L_3 \log (1+|z|^2)} e^{-\bar{z}L_+} |jj\rangle \ ,
\label{definition of coherent state 2}
\end{align}
where $z=\tan \frac{\theta}{2} e^{i\varphi}$.
An explicit form of $|\Omega\rangle$ is obtained from (\ref{definition of coherent
state 2}) as
\begin{align}
|\Omega\rangle=\sum_{m=-j}^{j}
\left( 
\begin{array}{c}
2j \\
j+m
\end{array}
\right)^{\frac{1}{2}}
\left( \cos \frac{\theta}{2} \right)^{j+m} \left( \sin \frac{\theta}{2} \right)^{j-m} 
e^{i(j-m) \varphi} |jm\rangle \ .
\label{explicit form}
\end{align}
It is easy to show the following relations by using (\ref{explicit form}):
\begin{align}
& \langle \Omega_1 | \Omega_2 \rangle
=\left( \cos\frac{\theta_1}{2}\cos\frac{\theta_2}{2}+e^{i(\varphi_2-\varphi_1)}
\sin\frac{\theta_1}{2}\sin\frac{\theta_2}{2} \right)^{2j} \ , 
\label{property2}\\
& |\langle \Omega_1 | \Omega_2 \rangle |= \left(\cos \frac{\chi}{2}\right)^{2j} 
\;\;\mbox{with} \;\;
\chi =\arccos (\vec{n}_1\cdot \vec{n}_2) \ , 
\label{property3}\\
& \frac{2j+1}{4\pi} \int d\Omega \ |\Omega\rangle\langle \Omega | =1 \ .
\label{property4}
\end{align}
Putting $\chi=\frac{2}{\sqrt{j}}$ in the the right-hand side of (\ref{property3}) gives rise to
\begin{align}
\left(\cos\frac{\chi}{2}\right)^{2j} \approx\left(1-\frac{1}{2j}\right)^{2j} \approx e^{-1} 
\end{align}
for large $j$.
This implies that the effective width of the Bloch coherent state is proportional to
$\frac{R}{\sqrt{j}}$.

\section*{Appendix B: The action with $\lambda=0$}
\setcounter{equation}{0}
\renewcommand{\theequation}{B.\arabic{equation}}
In this appendix, we describe how to calculate $F[x,\alpha=2]$ in the case of 
free fields. We extend the length of the time direction from $\beta$ to $2\beta$ and 
divide it into $2M$ sites, so that the lattice spacing $a$ is $a=\frac{\beta}{M}$.
We unify $\Phi_1$ and $\Phi_2$ into $\Phi(n)$ ($n=1,2,\cdots,2M$) such that
$\Phi(n)=\Phi_1(na)$ for $n=1,\cdots,M$ and $\Phi(n)=\Phi_2((n-M)a)$ for 
$n=M+1,\cdots,2M$.
Then, the discretized action with $\lambda=0$ is 
\beqa 
S_{NC}
&=&\frac{a}{2}\left[\sum_{m+m'\leq 2j-u}\left\{\sum_{n=1}^{M-1}\left|\frac{\Phi_{mm'}(n+1)-\Phi_{mm'}(n)}{a}\right|^2+\left|\frac{\Phi_{mm'}(1)-\Phi_{mm'}(M)}{a}\right|^2\right.\right.\nonumber\\
&\,&\,\,\,\left.+\sum_{n=M+1}^{2M-1}\left|\frac{\Phi_{mm'}(n+1)-\Phi_{mm'}(n)}{a}\right|^2+\left|\frac{\Phi_{mm'}(M+1)-\Phi_{mm'}(2M)}{a}\right|^2\right\}\nonumber\\
&\,&\,\,\,+\sum_{m+m'> 2j-u}\left\{\sum_{n=1}^{2M-1}\left|\frac{\Phi_{mm'}(n+1)-\Phi_{mm'}(n)}{a}\right|^2+\left|\frac{\Phi_{mm'}(1)-\Phi_{mm'}(2M)}{a}\right|^2\right\}\nonumber\\
&\,&\,\,\,\left.+\sum_{mm'}\sum_{n=1}^{2M}\left\{\Phi_{mm'}(n)\left[L_i,\left[L_i,\Phi(n)\right]\right]_{m'm}+m^2|\Phi_{mm'}(n)|^2\right\}\right] \ .
\label{discretized S}
\eeqa
Here we introduce a matrix $T_{nij,mkl}$ that is defined by
\beqa
S_{NC}=\sum_{n,l,m_1,m_2,m_3,m_4}\Phi^{*}_{m_1m_2}(n)
T_{nm_1m_2,lm_3m_4}\Phi_{m_3m_4}(l) \ .
\eeqa
We read off the matrix $T$ from (\ref{discretized S}) and calculate its determinant
numerically. Then, the free energy is given  by
\begin{align}
F[x,\alpha=2]=\frac{1}{2}\log\det T + \mbox{const.} \ .
\end{align}
The constant in the right-hand side does not contribute to the derivative of 
entanglement entropy with respect to $x$.

\end{document}